\documentclass[aps,prb,12pt]{revtex4}
\usepackage{graphicx}
\usepackage{subfigure}  % use for side-by-side figures
\begin{document}

\newcommand{\etal}{{\it et al.}\/}
\newcommand{\gtwid}{\mathrel{\raise.3ex\hbox{$>$\kern-.75em\lower1ex\hbox{$\sim$}}}}
\newcommand{\ltwid}{\mathrel{\raise.3ex\hbox{$<$\kern-.75em\lower1ex\hbox{$\sim$}}}}

\title{Doping Asymmetry of a 3-orbital CuO$_2$ Hubbard Model}

\author{Steven R. White}
\affiliation{University of California, Irvine,  Irvine, CA 92697, USA}
\author{D.J.~Scalapino}
\affiliation{Department of Physics, University of California, Santa Barbara, CA 93106-9530, USA}

%\date{\today}

\begin{abstract}
While both the hole and electron doped cuprates can exhibit $d_{x^2-y^2}$-wave
superconductivity, the local distribution of the doped carriers is known to be
significantly different with the doped holes going primarily on the O sites while
the doped electrons go on the Cu sites. Here we report the results of
density-matrix-renormalization-group calculations for a three-orbital
model of a CuO$_2$ lattice. In addition to the asymmetric dependence of the
intra-unit-cell occupation of the Cu and O for hole and electron doping, we
find important differences in the longer range spin and charge correlations.
As expected, the pair-field response has a $d_{x^2-y^2}$-like structure for both
the hole and electron doped systems.
\end{abstract}

%\pacs{71.10.Fd, 03.75.Ss, 74.25.Ha }

\maketitle
%\tableofcontents

%\section{Introduction}\label{sec:1}

How well does a 3-orbital Hubbard model describe the properties of the cuprates?
These materials are known to be charge-transfer systems and from the analysis of
Zaanan, Sawatzky and Allen \cite{ref:1} one would expect that a minimal model
which includes a Cu 3$d_{x^2-y^2}$ orbital and two O 2$p\sigma$ orbitals per unit cell would
be required. Indeed, early on a 3-orbital Hubbard model was proposed by several
groups \cite{ref:2,ref:3}, and various quantum Monte Carlo\cite{dopf,scalettar} and  embedded cluster 
calculations\cite{ref:17,ref:18} have shown
that this model exhibits a number of the basic magnetic and single particle spectral weight properties that are seen in the cuprates. More
recently, experimental measurements of both the hole and electron doped cuprates have provided
new information on the spatial charge and spin structure which can occur when
these materials are 
doped \cite{ref:4,ref:5,ref:6,ref:7,ref:8,ref:9,ref:10,ref:11,ref:12,ref:14a}.
So the question of whether a 3-orbital Hubbard model provides a suitable framework
with which to describe the physics of the cuprates has been enlarged. Here with the
experimental results for hole doped La$_{2-x}$Ba$_x$CuO$_4$ (LBCO) and electron
doped Nd$_{2-x}$Ce$_x$CuO$_4$ (NCCO) in mind, we have carried out density matrix
renormalization group (DMRG) \cite{ref:DMRG} calculations with the goal of determining
whether the 3-orbital Hubbard model remains an adequate model for the cuprates.

Neutron scattering studies of the LTT phase of LBCO find that the doped holes
form a striped structure consisting of regions with excess holes separated by
$\pi$-phase shifted antiferromagnetic regions \cite{ref:6}. At 1/8 hole doping,
superconducting correlations are observed to onset together with the stripe
order \cite{ref:7}. This pair-density-wave phase is believed to have a $d$-wave
pair-field which is large in the regions with excess holes and oscillates in
sign between these charged regions\cite{ref:8,ref:9}. Achkar et al. \cite{ref:10} have reported resonant
soft x-ray scattering measurements which show that the charge distribution on
the oxygens of LBCO have an $s'$-CDW orbital structure in which the charge modulations
on the O$_{p_x}$ and O$_{p_y}$ sites in a unit cell are in phase. 
STM Studies of BSCCO ($p\sim8$\%)
and NaCCOC ($p\sim12$\%) find that these materials have a predominantly $d$-CDW
orbital form factor in which these O$_{p_x}$ and O$_{p_y}$ charge modulations are out of 
phase \cite{ref:11}. Finally, recent resonant x-ray scattering
measurements of ${\rm Nd}_{2-x}{\rm Ce}_x{\rm CuO}_4$ near optimal doping \cite{ref:12}
find charge order which occurs with a similar periodicity and Cu-O bond orientation
to that of the charge stripes seen in LBCO.

One-band Hubbard and $t$--$J$ models have been found, within various approximations,
to exhibit striped charge and spin structures
\cite{ref:Poilblanc,ref:Zaanen,ref:Machida,ref:Schulz,ref:PRL80.1272,ref:PRB60.R753},
modulated nematic phases \cite{ref:Metlitski,ref:Sau,ref:Fischer} as well as pair density wave
phases \cite{ref:Corboz,ref:Berg,ref:Fradkin}. RPA calculations for the three-band
Hubbard model have also found nematic phases in certain parameter regimes
\cite{ref:Bulut,ref:Atkinson,ref:Maier}. Earlier DMRG calculations for
a 3-orbital model of a two-leg CuO$_2$ ladder showed the expected local asymmetric
charge-transfer behavior in which doped holes tend to predominantly go on the
$2p\sigma$ orbitals while doped electrons go on the Cu $3d_{x^2-y^2}$ orbitals
\cite{ref:jeckelmann,ref:nishimoto}. These calculations also found $d_{x^2-y^2}$-like pairing
correlations for both hole and electron doping in which the near neighbor Cu rung
and leg pair-field correlations differ in sign. Here we extend these calculations
to an $8\times4$  CuO$_2$ cluster with cylindrical boundary conditions.  The cylindrical boundaries reduce the edge effects associated with the ladder,  more reliably representing bulk behavior.  The $L\times4$ geometry is also the minimal size that can contain stripe-like clusters of holes.  With the $8\times4$ system we study the tendencies towards striping in the hole densities and whether doped holes or electrons modulate the phase of the antiferromagnetism. We also study the hopping kinetic energy associated with added holes or electrons, and the pairing tendencies in the doped system. 

%\section{The 3-orbital Model}\label{sec:2}

\begin{figure}[htbp]
\includegraphics[width=7.5cm]{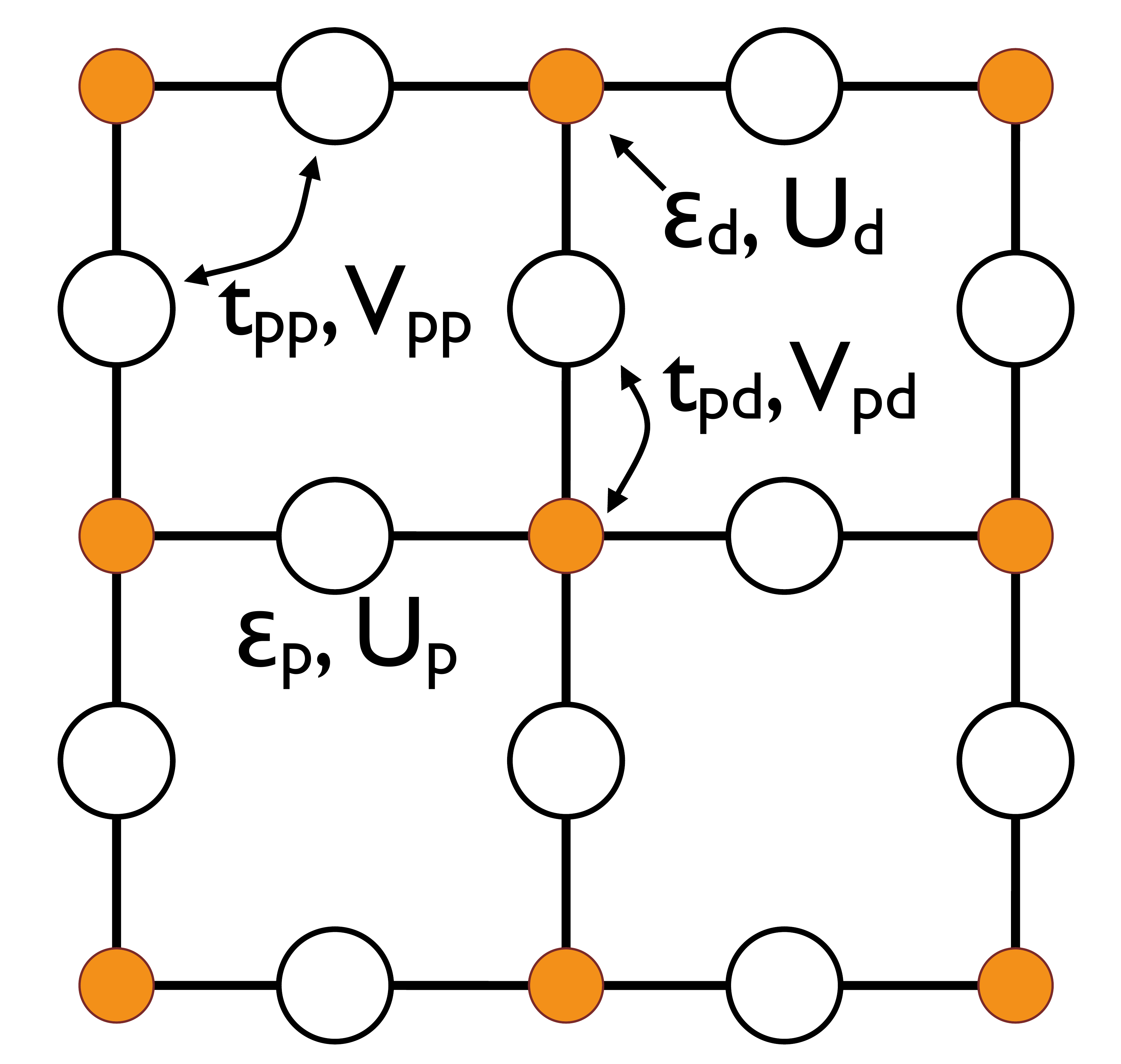}
%\framebox{\begin{minipage}{1in}\hfill\vspace{1in}\end{minipage}}
\caption{The CuO$_2$ lattice with onsite Cu $3d_{x^2-y^2}$ and O $2p_x$ and $2p_y$
energies $\epsilon_d$ and $\epsilon_p$, near neighbor Cu--O and O--O hoppings $t_{pd}$ and
$t_{pp}$, onsite Cu and O Coulomb interactions $U_d$ and $U_p$ and near neighbor
Cu--O and O--O Coulomb interactions $V_{pd}$ and $V_{pp}$, respectively. The
Hamiltonian for this 3-orbital model is given in a hole representation by
Eq.~(\protect\ref{eq:1}). Here we will work with energies measured in units of
$t_{pd}$, and take as a generic set of parameters $t_{pp}=0.5$,
$\Delta_{pd}=\epsilon_p-\epsilon_d=3$, $U_d=8$, $U_p=3$, $V_{pd}=1$ and $V_{pp}=0.75$.
\label{fig:1}}
\end{figure}

The lattice structure and the parameters of the three orbital CuO$_2$ model
that we will study are shown in Fig.~\ref{fig:1}.
%This is code for a sample figure:
The model has a CuO$_2$ unit cell consisting of a $3d_{x^2-y^2}$ orbital on the
Cu site and $2p_x/2p_y$ orbitals on the $x$ and $y$ oxygens. In a representation
in which the vacuum of the 3-orbital model has the configuration ($d^2_{x^2-y^2}p^2_xp^2_y$),
$d^+_{i\sigma}$ and $p^+_{j\sigma}$ create holes with spin $\sigma$ on the
$i^{th}$ Cu and $j^{th}$ O sites respectively, and the Hamiltonian has the form
\begin{eqnarray}
  H&=&\Delta_{pd}\sum_{i\sigma}p^+_{i\sigma}p_{i\sigma}-t_{pd}\sum_{\langle ij\rangle\sigma}
	(d^+_{i\sigma}p_{j\sigma}+p^+_{j\sigma}d_{i\sigma})\cr
	&-&t_{pp}\sum_{\langle ij\rangle\sigma}(p^+_{i\sigma}p_{j\sigma}+p^+_{j\sigma}p_{i\sigma})\cr
	&+&U_d\sum_in^d_{i\uparrow}n^d_{i\downarrow}+U_p\sum_in^p_{i\uparrow}n^p_{i\downarrow}\cr
	&+&V_{pd}\sum_{\langle ij\rangle}n^d_in^p_j+V_{pp}\sum_{\langle ij\rangle}n^p_in^p_j
\label{eq:1}
\end{eqnarray}
Here $\Delta_{ps}=\epsilon_p-\epsilon_d$ is the energy difference between
having a hole on an O site versus a Cu site, $t_{pd}$ and $t_{pp}$ are one-hole
hopping matrix elements between near-neighbor Cu and O sites and near-neighbor
O sites, respectively. The sums $\langle ij\rangle$ in Eq.~\ref{eq:1} denote
sums over the relevant nearest-neighbor sites. $U_d$ and $U_p$ are the onsite Cu
and O Coulomb interactions and $V_{pd}$ and $V_{pp}$ are the nearest-neighbor
Cu--O and O--O Coulomb interactions, respectively. The phases of the orbitals
have been fixed such that the signs of the hopping matrix elements remain the
same throughout the lattice and are positive. 

The hopping parameters $t_{pd}$
and $t_{pp}$ found for La$_2$CuO$_4$ and Nb$_2$CuO$_4$ in various cluster and
LDA calculations are relatively close to each other. We will measure energies
in units of $t_{pd}$ and set $t_{pp}/t_{pd}=0.5$ for both of these
materials \cite{ref:5,ref:15}. The primary difference in the one-electron
parameters occurs in $\Delta_{pd}$ where the absence of the apical oxygens in
Nd$_2$CuO$_4$ is expected to lead to a reduction in $\Delta_{pd}$ relative to
La$_2$CuO$_4$. Indeed this is found in LDA calculations, however the appropriate
bare values of $\Delta_{pd}$ to use in the 3-orbital Hamiltonian Eq.~(\ref{eq:1})
has posed a problem because of double counting corrections \cite{ref:16,ref:17}.
Here we find that setting $\Delta_{pd}/t_{pd}=3$ gives reasonable values for the
charge gap and exchange interaction. Thus, working in units of $t_{pd}$ we will
take for a canonical set of parameters
\begin{equation}
  t_{pp}=0.5,\hspace{0.5mm}\Delta_{pd}=\varepsilon_p-\varepsilon_d=3,\hspace{0.75mm}
  U_d=8,\hspace{0.75mm}U_p=3,\hspace{0.75mm}V_{pd}=1,\hspace{0.75mm}V_{pp}=0.75
\label{eq:2}
\end{equation}
These parameters are appropriate for a charge transfer system
for which $U_d>\varepsilon_p-\varepsilon_d=\Delta_{pd}$ and $\Delta_{pd}>2t_{pd}$.
Using a similar set of parameters for a 2-leg CuO$_2$ ladder we previously found
at half-filling a charge gap $\Delta_c\sim t_{pd}$ and a spin gap $\Delta_s\sim0.03
t_{pd}$. For a 2-leg ladder the effective exchange coupling $J\sim2\Delta_s\sim0.06
t_{pd}$. For $t_{pd}$ of order 1 to 2 eV, these correspond to reasonable values
for the charge gap and the exchange interaction. Our plan is to use this same
set of parameters for both the hole and electron doped systems and focus on the
differences that arise between them. We will comment on the effect of reducing
$t_{pp}$ and, for the electron doped case, the effect of reducing $\Delta_{pd}$.

The DMRG calculations will be carried out for an $8\times4$ CuO$_2$ lattice which
has periodic boundary conditions in the 4-unit cell $y$-direction and open ends
in the 8-unit cell $x$-direction. For the charge and spin studies, we will work with a fixed number of holes
$32+N$ and a hole density per CuO$_2$ unit cell $x=1+N/32$ which is 1 for the undoped system. Positive
values of $N$ ($x>1$) correspond to hole doping and negative values of $N$
($x<1$) to electron doping. We typically did 15 DMRG sweeps, keeping up to $m=4000$ states on the last sweep.
This led to excellent convergence for the local quantities that we report here.
A typical maximum truncation error was $\sim 10^{-5}$; extrapolating the truncation error to zero gave
typical fractional errors in the total energy also about $\sim 10^{-5}$. Without extrapolation,
fractional errors in energy were estimated to be less than $10^{-4}$, 
and absolute errors in local quantities were in the range $10^{-3}-10^{-4}$.  
The good overall convergence for this cluster suggests that wider systems, say up to width $6$, 
will be accessible for near-future studies.

\begin{figure}[htbp]
\includegraphics[width=12cm]{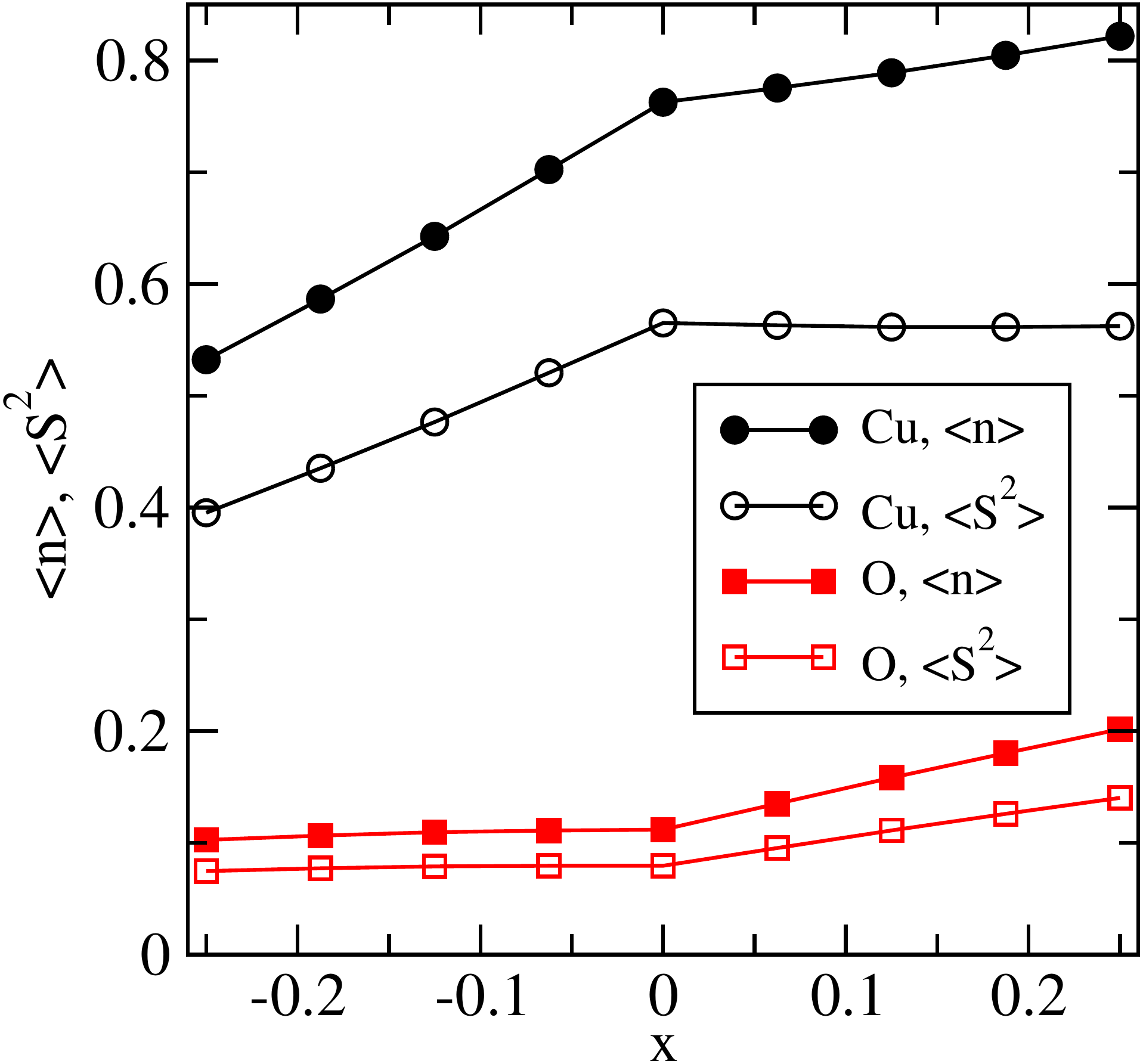}
%\framebox{\begin{minipage}{1in}\hfill\vspace{1in}\end{minipage}}
\caption{The hole density $\langle n\rangle$ and the square of the spin moment
$\langle S^2\rangle = \frac{3}{4}\langle n_\uparrow+n_\downarrow-2n_\uparrow
n_\downarrow\rangle$ on the Cu (black circles) and O (red squares) sites versus
the hole density $x=1+N/32$ per CuO$_2$ unit cell. The undoped $8\times4$ CuO$_2$
lattice has 32 holes and $x=1$. $N > 0$ ($x > 1$) corresponds to doping additional
holes while $N < 0$ ($x < 1)$ corresponds to electron doping which reduces the
number of holes.
\label{fig:2}}
\end{figure}

In Fig.~\ref{fig:2} we show the effect of doping on the local charge density and
squared spin moments on the Cu and O sites as a function of the hole density
$x$. As we will discuss later, there can be inter- and
intra-cell spatial structure in the charge and spin. The results shown in
Fig.~\ref{fig:2} represent site averages taken over the $8\times4$ lattice.
For the undoped $x=1$ ($N=0$) system where there is one hole per CuO$_2$ unit
cell, Fig.~\ref{fig:2} shows that the hole occupation is approximately 80\%
on the Cu site and 10\% on each of the two O sites for the parameters that we
have chosen. When additional holes are added they go approximately 75\% onto
the two O sites and 25\% onto the Cu site of the unit cell. Alternatively,
under electron doping, the added electrons go approximately 90\% onto the Cu site
and only 10\% onto the two O sites. This is of course what one would expect for
a charge-transfer system. The change in the square of the spin moments on the Cu
and O sites is seen to vary with the hole concentration $x$ in a similar manner
to that of the charge occupation. For electron doping ($x<1$), an electron added
to a Cu site removes the hole spin moment leading to a decrease in
$\langle S^2\rangle$ averaged over the lattice, while for hole doping the square
of the O hole spin moment increases as holes are primarily added to the O sites.

\begin{figure}[htbp]
\includegraphics[width=14cm]{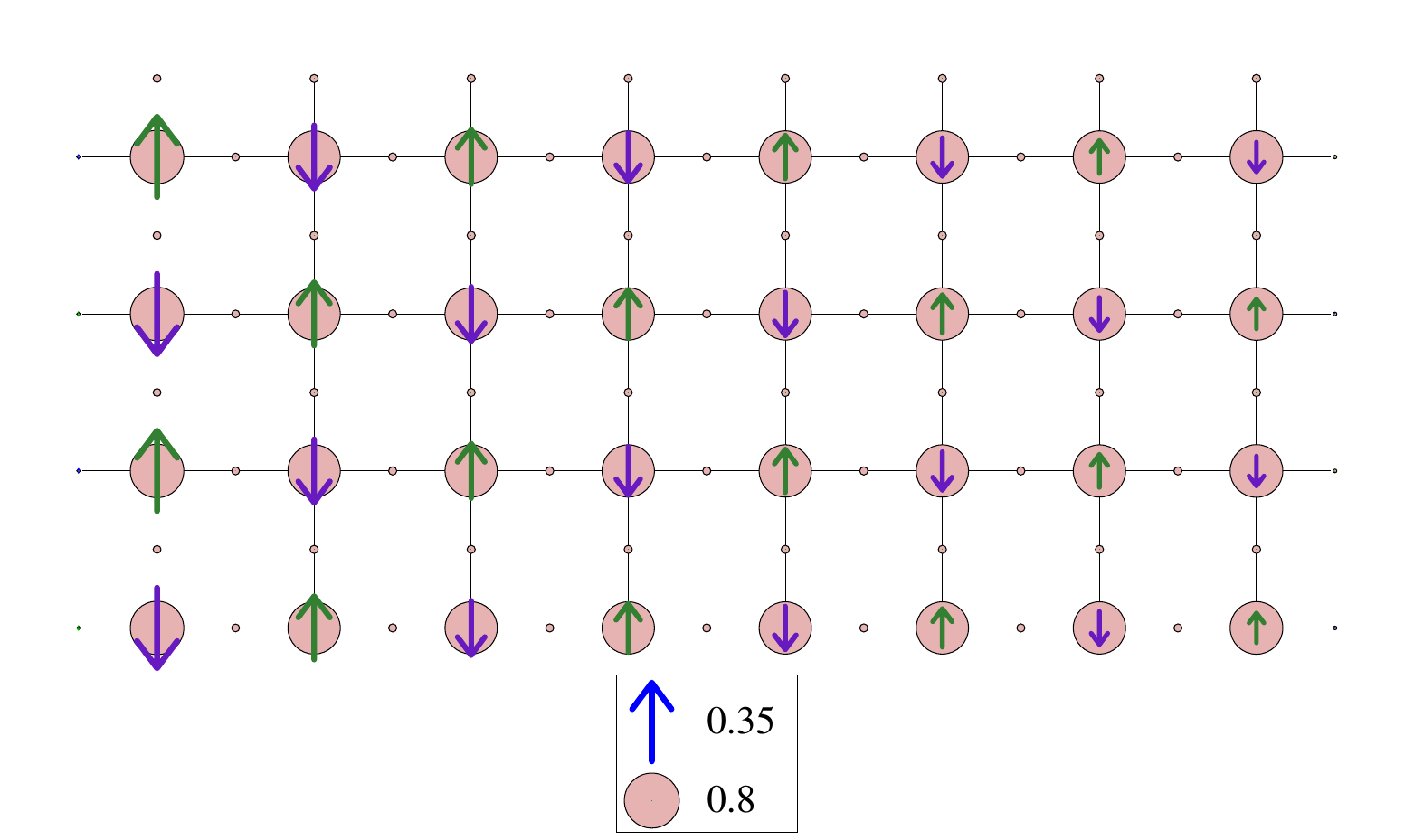}
%\framebox{\begin{minipage}{1in}\hfill\vspace{1in}\end{minipage}}
\caption{The hole occupation $\langle n \rangle$ and spin structure $\langle S_z \rangle$ 
for the undoped (32 hole)
$8\times4$ CuO$_2$ lattice. The hole occupation is proportional to the 
diameter of the circles.
A staggered magnetic field of
magnitude $h=0.1$ was applied to the Cu sites along the left-hand edge of the
$8\times4$ CuO$_2$ lattice which has periodic boundary conditions in the $y$
direction and open end boundary conditions in the $x$-direction.
\label{fig:3}}
\end{figure}

To study the longer range spin and charge correlations, we have applied a weak
staggered magnetic field to the Cu sites on the left hand edge of the $8\times4$
lattice. The expected antiferromagnetic response of the undoped system is shown
in Fig.~\ref{fig:3}. Here, the diameters of the circles are proportional to the density
of the holes and the lengths of the arrows are proportional to the spin moments.
One sees, as shown in Fig.~\ref{fig:2} that the holes are mainly on the Cu sites.
The applied edge field has broken the spin symmetry and there is a well formed
antiferromagnetic spin pattern.

\begin{figure}[htbp]
\includegraphics[width=12cm]{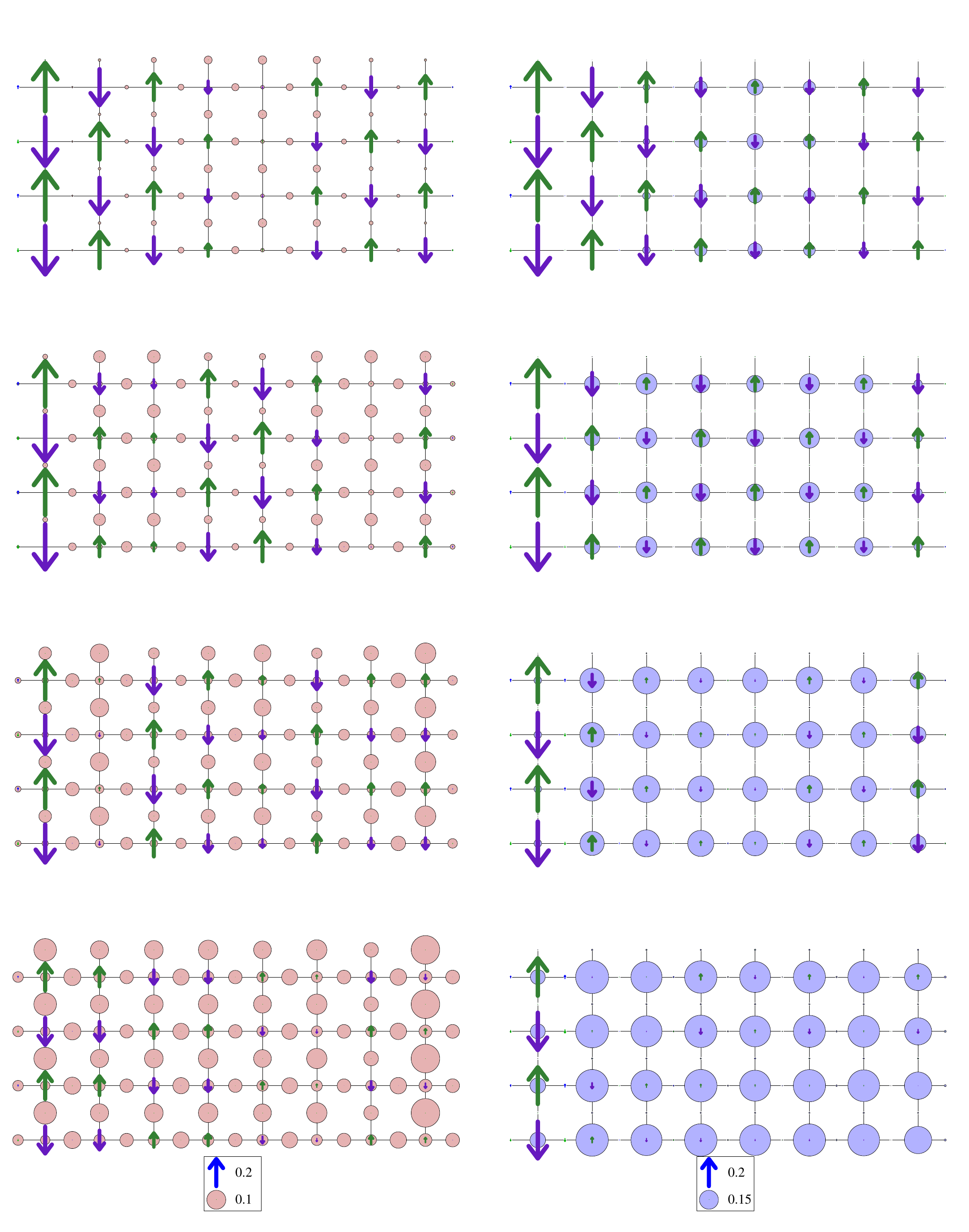}
%\framebox{\begin{minipage}{1in}\hfill\vspace{1in}\end{minipage}}
\caption{Charge and spin structure for the hole (left) and electron (right)
doped lattices for the parameters given in Eq.~(\protect{\ref{eq:2}}). Here the
hole density distribution for the undoped lattice shown
in Fig.~\protect\ref{fig:3} has been subtracted. The added hole density on a site
is proportional to the diameter of the red circles shown on the left. Similarly,
the diameter of the blue circles on the right is proportional to the added
electron density. Note the difference in the diameter scales for the hole and
electron doped figures. The small $\sim5$\% of the additional electron density that
goes onto an O site is not visible on this scale. For the $8\times4$ lattice, the left hand figures (top to bottom)
correspond to the addition of 2,4,6 and 8 holes respectively, while the right hand
figures correspond to the addition of a corresponding number of electrons. A
staggered magnetic field of magnitude $h=0.1$ has been applied to the Cu sites
on the left-hand edge of the lattice.
%Indications
%of striping are denoted by marks below some of the lattices.
\label{fig:4}}
\end{figure}

In Fig.~\ref{fig:4} we contrast the results for hole doping on the left with
electron doping on the right. In this figure, the hole density distribution of
the undoped lattice shown in Fig.~\ref{fig:3} has been subtracted.
The diameters
of the circles for hole doping on the left are proportional to the added hole density
while the diameters of the circles on the right are proportional to the added
electron density. In this figure the diameter scale used for the hole density is
0.1 and for the electron density 0.15. In the top left hand lattice shown in Fig.~\ref{fig:4}, 2 holes
have been added to the 32 holes of the undoped $8\times4$ lattice giving $x=1.0625$.
The lattices shown below this have 4,6 and 8 holes added corresponding to hole
concentrations $x$ per CuO$_2$ unit cell of 1.125, 1.1875 and 1.25, respectively.
The lattices on the right hand side of Fig.~\ref{fig:4} show similar results for
the case in which electrons are added (or holes removed). From top to bottom
these lattices have 30, 28, 26 and 24 holes, respectively, corresponding to
$x=0.9375$, 0.875, 0.8125 and 0.75. As in Fig.~\ref{fig:3}, a staggered magnetic field
($h=0.1$) was applied to the Cu sites on the left-hand edge of the lattice.

\begin{figure}[htbp]
\includegraphics[width=7.5cm]{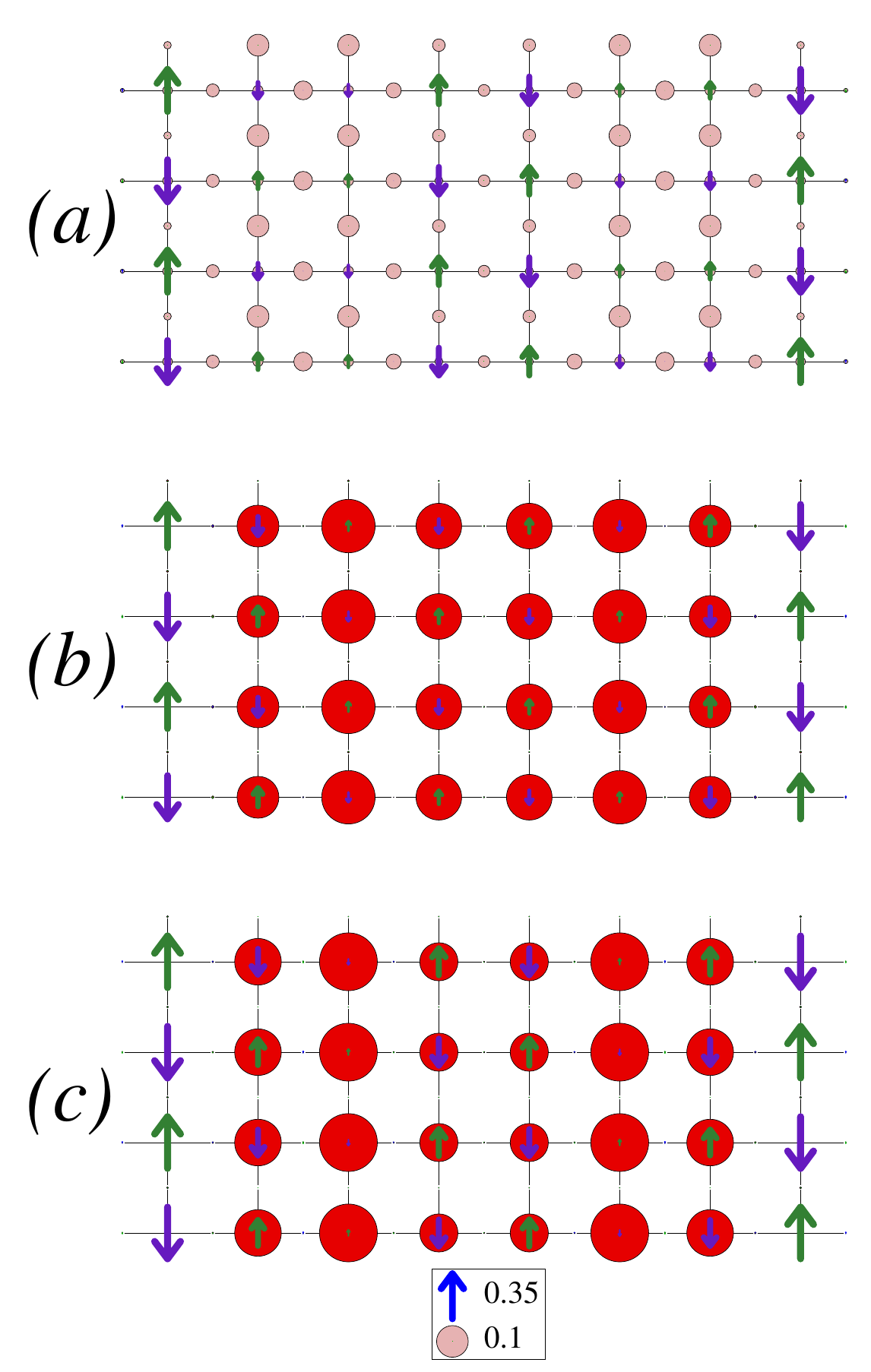}
%\framebox{\begin{minipage}{1in}\hfill\vspace{1in}\end{minipage}}
\caption{(a) The charge and spin structure of the 1/8 ($x=1.125$) hole doped
system with a weak staggered magnetic field $h=0.1$ applied to both ends of the
$8\times4$ lattice. Here an $s'$-CDW--SDW structure is seen.
(b) The charge and spin structure of the 1/8 ($x=0.875$) electron doped system
with $t_{pp}=0.5$. Here we find only a weak charge modulation and a commensurate spin
antiferromagnetic structure.
(c) Similar to the $x=0.875$ electron doped system shown in (b) but with
$t_{pp}=0.0$. In this case there is an incommensurate antiferromagnetic spin
structure similar to that of the hole doped system.
\label{fig:5}}
\end{figure}

As shown in Fig.~\ref{fig:2}, the additional holes tend on average to go onto
the O sites, but as seen in Fig.~\ref{fig:4} their distribution is not uniform.
For hole doping there is a tendency for stripe formation separated by $\pi$-phase
shifts in the antiferromagnetic correlations. For $x=1.125$ there are two, approximately
Cu-site centered, stripes separated by a $\pi$-phase shifted antiferromagnetic
region similar to the well known behavior of La$_{1.48}$Nd$_{0.4}$Sr$_{0.12}$CuO$_4$
\cite{ref:6}. For $x=1.0625$ there is a single stripe and for $x=1.1875$, corresponding to the
addition of 6 holes on the $8\times4$ lattice, one can see the remnants of a
three stripe structure. This structure vanishes for the strongly overdoped
$x=1.25$ case. The stripe spacing for the three lower hole dopings is consistent
with the relation $d^{-1}=2(x-1)$ and the well known spin $\delta_{\rm spin}$
and charge $\delta_{\rm charge}$ incommensurability relation
$\delta_{\rm charge}=2\delta_{\rm spin}$ found in the La-based cuprates \cite{ref:4,ref:6}.
A closer look at the structure of the charge and spin distributions for the 1/8 ($x=1.125$)
hole doped lattice is shown in Fig.~\ref{fig:5}(a). Here a weak staggered
magnetic field has been applied to both ends of the $8\times4$ lattice. In this
case, the $x=1.125$ hole doped system exhibits bond centered charged stripes
separated by $\pi$-phase shifted antiferromagnetic regions. The charge modulations
on the $O_{p_x}$ and $O_{p_y}$ sites are in phase leading to what was called an
$s'$--CDW--SDW phase in Ref.~\onlinecite{ref:10}. There may be an additional small admixture
of $d$-CDW. Of course the $8\times4$ lattice already breaks C4 symmetry so one
expects differences in the $x$ and $y$ oxygen hole occupations. Increasing
$V_{pp}$ leads to an increase in these differences \cite{ref:KFG} but the $s'$ symmetry
remains dominant.

For the electron doped system, one sees on the right hand side of
Fig.~\ref{fig:4} that the spin and charge structure appears quite different from
the hole doped case. Of course the doped electrons go dominantly on to the Cu
sites and initially the small concentration of added electrons are repelled from
the open edge boundaries by the ``infinite" edge potential. However, for these
parameters, by the time the electron doping reaches 0.125, a relatively uniform
density of the added electrons is spread over the Cu sites and the
antiferromagnetic order remains. A closer look at the 1/8 electron doped lattice
is shown in Fig.~\ref{fig:5}(b). Here one can see that there are two charge stripes
but the antiferromagnetic correlations remain commensurate. Thus for these
parameters we find charge stripes with incommensurate antiferromagnetism for
hole doping and commensurate antiferromagnetic spin correlations for electron
doping. This remains the case for the electron doped system when $\Delta_{pd}$
is reduced as is expected in the $T'$ structure where the apical oxygens are absent.
Another important parameter is $t_{pp}$ which determines the effective hopping
$t'$ between next-near neighbor Cu sites. In Hubbard and $t-t'-J$ models it is
known that $t'$ affects the stripe stability \cite{ref:8,ref:PRB60.R753}.
Here we find that when the oxygen-oxygen hopping $t_{pp}$ is reduced, the
amplitude of the charge stripes is increased and the spin
structure for the electron doped system also becomes incommensurate as shown in
Fig.~\ref{fig:5}(c) for $t_{pp} = 0$. The effect of reducing $t_{pp}$ acts to
increase the frustration associated with the antiferromagnetic background and
gives rise to the $\pi$-phase shifted antiferromagnetic regions separating the
charge stripes.
We find that when $t_{pp}$ is reduced (below $\ltwid0.25$), striping can occur for both
the electron and hole doped system. However, the tendency for striping is
stronger in the hole doped system.

\begin{figure}[htbp]
\includegraphics[width=12cm]{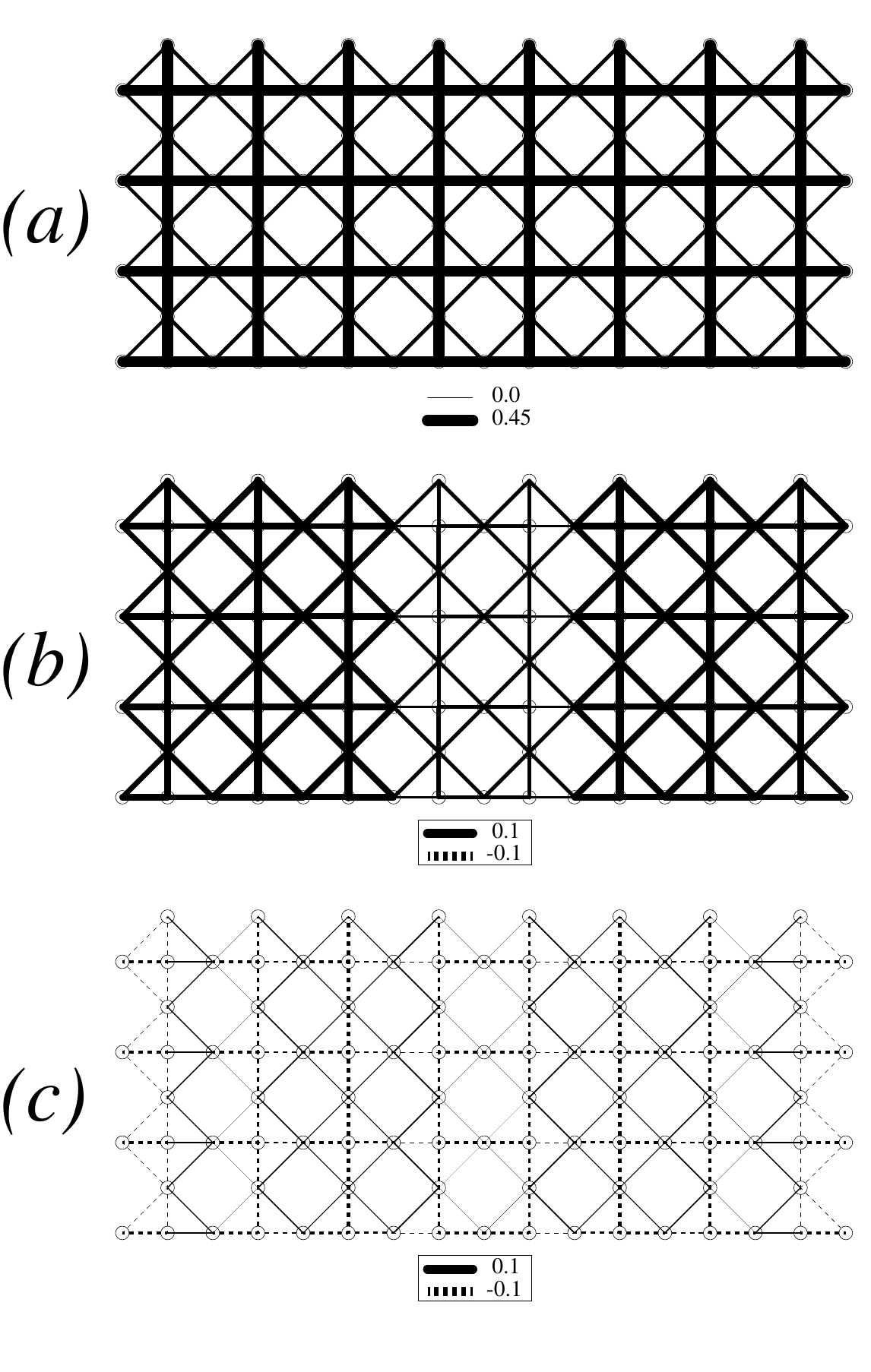}
%\framebox{\begin{minipage}{1in}\hfill\vspace{1in}\end{minipage}}
\caption{(a) The hopping strength on each bond for the undoped system.  (b) The difference in the hopping 
strength relative to the undoped system for the
hole doped system with $x = 1.125$. (c) The same as (b), but for 
the electron doped system,  $x = 0.875$ 
\label{fig:6}}
\end{figure}

The addition of holes to the filled band vacuum configuration
($d^2_{x^2-y^2}p^2_xp^2_y$) lowers the kinetic energy. For the hole Hamiltonian,
Eq.~(\ref{eq:1}), with positive hopping parameters this means that one expects
the Cu-O and O-O hole hopping strengths $\sum_s\langle d^+_{is}p_{js}+p^+_{js}d_{is}\rangle$
and $\sum_s\langle p^+_{is}p_{js}+p^+_{js}p_{is}\rangle$ to be positive for the
32 hole doped system as illustrated in Fig.~\ref{fig:6}(a).

The Cu-O hopping strength is larger than the O-O hopping strength reflecting the
fact that the doped holes are of order 80\% on the Cu sites. When additional
holes are added, the hopping strength increases further. The difference in the
hopping strengths between the 36 hole doped lattice and the undoped 32 hole lattice
are illustrated in Fig.~\ref{fig:6}(b). In the case of hole doping, the holes are
distributed to both the Cu and O sites ($\sim25$\% to the Cu and $\sim37.5$\% to
each of the O sites) leading to the enhancement of both the Cu-O and O-O hopping
strengths shown in Fig.~\ref{fig:6}(b). In addition, one sees evidence of the
charge stripe structure.

For the case of electron doping, the electrons go dominantly on the Cu sites.
This reduction of the average number of holes on the Cu sites leads to the
reduction in the Cu-O hole hopping strength as shown in Fig.~\ref{fig:6}(c).
Although the reduction of the average hole occupation on each O is only of
order 5\%, one might have expected that this would also reduce the strength of
the O-O hole hopping. However, as shown in Fig.~\ref{fig:6}(c), the O-O hopping
strength is in fact slightly increased.
%We believe that this is an effect of
%the Coulomb correlations. A fluctuation in which a given Cu site has an additional
%electron will tend to have surrounding O sites with excess holes leading to an
%enhancement of the average O-O hole hopping. 
The overall change in
hopping strength is significantly smaller for the electron doped system.  The
total change in the kinetic energy measured in units of $t_{pd}$ per added hole
is of order $-3.2$ while per added electron it is only $+0.7$. This is consistent
with the notion that the doped holes will enter a region of the band between
$\Gamma$ and $M$ where there is significant dispersion while the electrons will
enter near $X$ where the dispersion is flat. 

\begin{figure}[htbp]
\includegraphics[width=12cm]{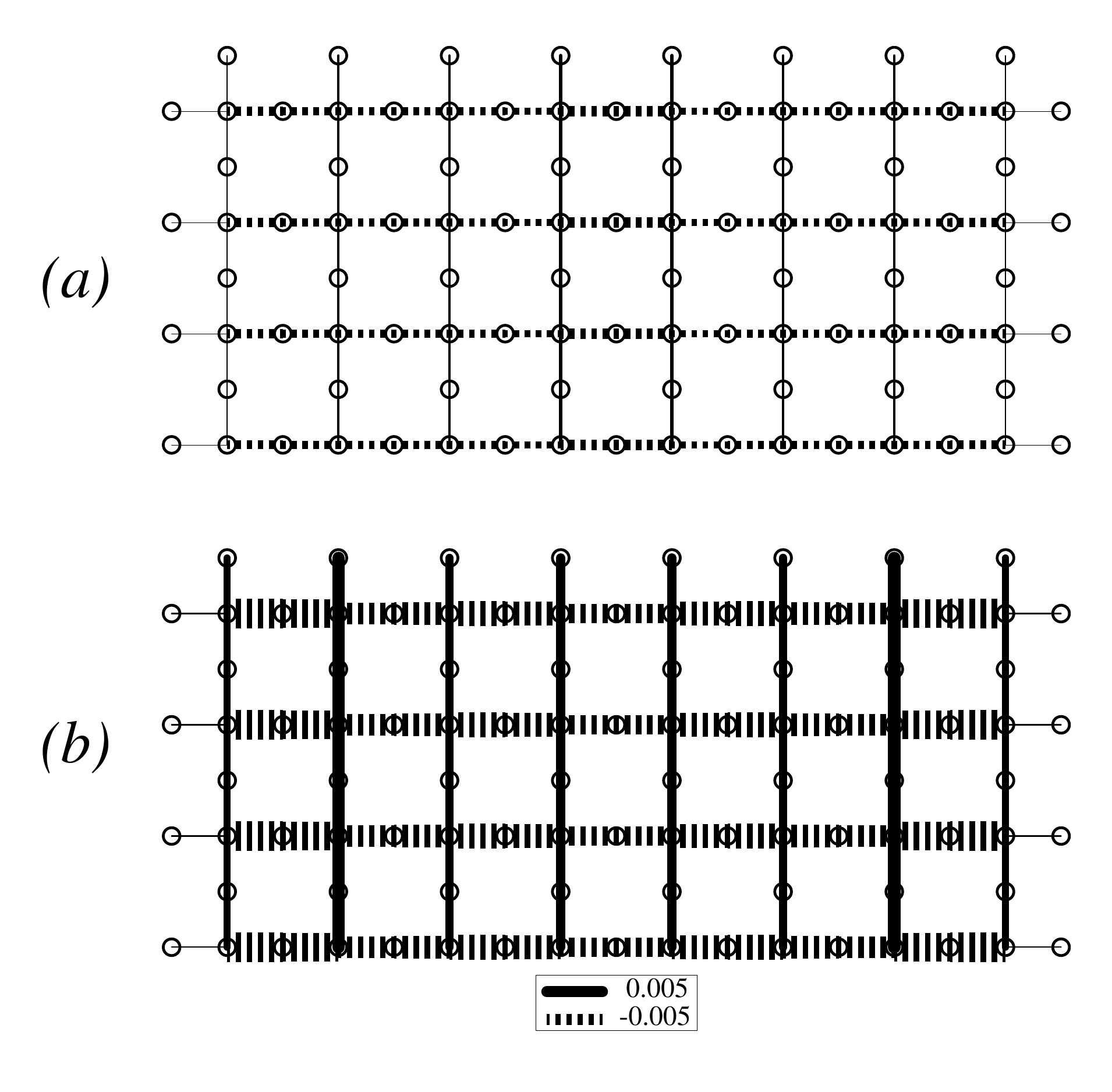}
%\framebox{\begin{minipage}{1in}\hfill\vspace{1in}\end{minipage}}
\caption{The pair-field induced by the proximity pair-field, Eq. (3), for the
hole $x \approx 1.125$ (a) and electron $x \approx 0.875$ (b) doped systems. Here the dashed
lines denote a negative pair-field amplitude between near neighbor Cu-O pairs of sites
while the solid lines denote a positive pair-field amplitude. The applied
proximity pair-field is between near neighbor Cu-Cu pairs of sites in the
x-direction with magnitude $\Delta_0=0.2$, and the response shows the expected $d_{x^2-y^2}$-like behavior for both the hole and
electron doped systems.
\label{fig:7}}
\end{figure}

In order to study the pairing response, we have applied a proximity singlet
pair-field that couples to near neighbor Cu sites along the $x$ direction,
\begin{equation}
\frac{\Delta_0}{2}\sum_{(\ell_x,\ell_y)}\left(\Delta^+_x(\ell_x,\ell_y)+\Delta_x(\ell_x,\ell_y)\right)
\label{eq:3}
\end{equation}
with $\Delta_x(\ell_x,\ell_y)=\left(d_{\ell_x+1,\ell_y\uparrow}
d_{\ell_x,\ell_y\downarrow}-d_{\ell_x+1,\ell_y\downarrow}d_{\ell_x,\ell_y\uparrow}\right)/\sqrt{2}$.
The Cu-O near neighbor responses in the $x$-direction $\langle\Delta_x(\ell_x,\ell_y)+\Delta_x^+(\ell_x,\ell_y)\rangle/2$
has a negative sign and is shown as the dashed lines in Fig.~\ref{fig:7}.
The solid lines, which indicate a positive value, show the pair-field response
$\langle\Delta_y(\ell_x,\ell_y)+\Delta_y^+(\ell_x,\ell_y)\rangle/2$ between near neighbor Cu-O sites in
the $y$ direction. 
For these calculations the
average hole number was set by a chemical potential $\mu$. In the top panel,
Fig. 7(a), $\mu=-1.0$, giving an average hole number $\langle N\rangle =36.05$ 
($x \sim 1.125$) while for
the lower panel, Fig. 7(b), $\mu=-3.2$ giving $\langle N\rangle  = 28.14$ ($x \sim 0.875$). 
Both the hole
doped pair-field response shown in Fig. 7(a) and the electron doped case shown in
Fig. 7(b) have the expected d-wave-like sign
change. 

If the proximity pair-field is applied only between the horizontal Cu-Cu sites on the left 
edge of the lattice, the induced pair-field decays rapidly in the x-direction for the hole doped system and somewhat more slowly for the electron doped case.  The longer range pair-field correlations are suppressed by finite size effects.   These are particularly severe for the periodic
in $y$ (tube-like) geometry of our CuO$_2$ lattice. As seen in Fig.~\ref{fig:4} for
the hole doped lattice, a stripe appears each time a pair of holes is added for
2, 4 and 6 holes. This is consistent with previous 2-leg ladder studies where it
was found that the preferred filling was 2 holes per 4 rungs \cite{ref:PRL80.1272,ref:PRB60.R753}.
Thus for an $8\times4$ CuO$_2$ tube a low-energy fluctuation of $\pm2$ holes
involves the creation or destruction of a stripe, 
leading to a high energy spin configuration
with a domain wall without holes.
Alternatively, one could consider a configuration which has 4 holes in a
stripe, but this is also energetically unfavorable. While this effect is less severe
for the electron doped system shown in Fig. 7(b) and the pair-field response is
stronger because it lacks the antiferromagnetic domain walls, we expect that 
the small $8\times 4$ size of
the lattice still acts to suppress the hole number fluctuations. 
Properly comparing the pairing between the electron doped and hole doped systems will require
larger systems, and if there are stripes they should be long, running lengthwise down the 
cylinder, either horizontally or spiraling.
%Thus states
%consisting of a superposition of
%configurations with different numbers of pairs are suppressed by the finite size and geometry
%of this CuO2 lattice.

In summary, we have studied an $8\times4$ three orbital Hubbard model for CuO$_2$
with parameters chosen to give a realistic charge gap and exchange coupling. With
one hole per CuO$_2$ unit, the hole occupation is approximately 80\% on the Cu
$d_{x^2-y^2}$ orbital and in the presence of a weak staggered edge magnetic field
commensurate antiferromagnetic correlations are found to extend across the lattice.
When additional holes are added they go $\sim75$\% onto the O sites and charge
stripes separated by $\pi$-phase shifted antiferromagnetic regions appear. The O
hole occupation and the Cu spin structure has an $s'$-CDW-SDW like structure \cite{ref:10}.
When additional electrons are added, they go approximately 90\% onto the Cu $d_{x^2-y^2}$
orbitals. For our small cluster, there is a weak tendency for charge modulations
but the antiferromagnetic spin correlations remain commensurate.
However, when the oxygen-oxygen one electron hopping $t_{pp}$ is reduced, a clear striped structure appears with
incommensurate antiferromagnetic correlations. For both the hole and
electron doped systems, the response of the $y$-near-neighbor Cu-Cu pair-field is
out of phase ($d$-wave like) with respect to the $x$-near neighbor Cu-Cu
pair-field induced by an applied $x$-near-neighbor proximity
pair-field. These pair-field correlations are short range reflecting the finite size
and geometric restrictions of the CuO$_2$ cluster studied.

%\section*{Acknowledgments}

We would like to thank J.C.S.~Davis, S.A.~Kivelson, M.A.~Metlitski, S.~Sachdev and
J.M.~Tranquada for insightful discussions. SRW acknowledges support from
the NSF under grant DMR-1161348 and from the Simons Foundation through the Many Electron collaboration. DJS acknowledges the support of the Center for Nanophase Materials Science at
ORNL, which is sponsored by the Division of Scientific User Facilities, U.S. DOE.

%This is a sample list of references:

\end{document}